\documentclass[prb,aps,amssymb,twocolumn,showpacs,amsmath]{revtex4}

\usepackage{graphicx}

\begin{document}

\title{Molecular spintronics: Coherent spin transfer in coupled quantum dots}
\author{Florian Meier$^{1,2}$}
\email{florian@iquest.ucsb.edu}
\author{Veronica Cerletti$^2$}
\author{Oliver Gywat$^2$}
\author{Daniel Loss$^2$}
\author{D.~D. Awschalom$^{1,3}$}
\affiliation{$^1$California Nanosystems Institute, University of
California, Santa Barbara, CA 93106, USA
\\ $^2$Department of Physics and Astronomy, University
of Basel, Klingelbergstrasse 82, 4056 Basel, Switzerland \\
$^3$Department of Physics, University of California, Santa
Barbara, CA 93106, USA}
\date{\today}

\begin{abstract}
Time-resolved Faraday rotation has recently demonstrated coherent
transfer of electron spin between quantum dots coupled by
conjugated molecules. Using a transfer Hamiltonian ansatz for the
coupled quantum dots, we calculate the Faraday rotation signal as
a function of the probe frequency in a pump-probe setup using
neutral quantum dots. Additionally, we study the signal of one
spin-polarized excess electron in the coupled dots. We show that,
in both cases, the Faraday rotation angle is determined by the
spin transfer probabilities and the Heisenberg spin exchange
energy. By comparison of our results with experimental data, we
find that the transfer matrix element for electrons in the
conduction band is of order $0.08 \,{\rm eV}$ and the spinFaraday
rotation signal as a function of the probe frequency in a
pump-probe setup transfer probabilities are of order $10$\%.
\end{abstract}


\maketitle

\section{Introduction}
\label{sec:intro}

The past years have evidenced rapid experimental progress in the
field of spintronics.~\cite{wolf:01,awschalom:02} Coherent
transport of electron spins in semiconductors has been
demonstrated over several micrometers,~\cite{kikkawa:97}
nourishing hopes that the electron spin may be used as carrier of
information similar to the electron charge. Such applications of
the spin degree of freedom for classical or quantum information
processing~\cite{loss:98} require control of the electron spin not
only in extended systems such as two-dimensional electron gases
(2DEG's), but rather also for spins localized in quantum dots
(QD's).

Recently, coherent transfer of electron spin has been observed
between QD's with different radii $r_A \simeq 1.7\,{\rm nm}$ (QD
$A$) and $r_B \simeq 3.5\,{\rm nm}$ (QD $B$) coupled by a benzene
ring.~\cite{ouyang:03} The different QD radii give rise to
different quantum size levels for electrons and holes in the two
species of QD's. The resulting difference in exciton energies
allows one to pump and probe selectively the spin polarization for
QD's of species $A$ and $B$. The main result of
Ref.~\onlinecite{ouyang:03} is that an electron spin polarization
created by optical pumping in QD $B$ is transferred
`instantaneously' to QD $A$. The efficiency of this transfer
mechanism is of order $10$\% at low temperatures $T<50\,{\rm K}$
and increases to approximately $20$\% for $T \gtrsim 100  \, {\rm
K}$. The observed shift of the exciton energies to lower values
compared to isolated QD's is also consistent with a coherent
delocalization of the electron or hole over the system formed by
the QD's and the bridging molecule.

The purpose of this paper is to show that a two-site Hamiltonian
with a transfer term captures some of the essential experimental
features. We aim at calculating the dependence of the
experimentally observed Faraday rotation (FR) signal as a function
of probe energy on microscopic parameters such as spin transfer
probabilities. The FR angle is proportional to the difference in
refractive indices for $\sigma^\pm$ circularly polarized light
which is determined by the difference of the dielectric response
functions. We calculate the dielectric response functions of
coupled QD's and derive an analytical expression for the FR angle
in terms of electron transfer probabilities and Heisenberg
exchange splittings. The experimental data provide strong evidence
that the spin transfer is mediated by the $\pi$-conjugated
molecule.~\cite{rem1} We do not aim to describe this transfer
mechanism microscopically, but consider the transfer matrix
elements for electrons and holes as parameters of the Hamiltonian.

For CdSe QD's with radii $r_A$ and $r_B$, the single-particle
level spacing for electrons and holes is large compared to the
temperatures $T \leq 200 \, {\rm K}$ explored experimentally. This
allows us to restrict our attention to the lowest orbital levels
in the conduction and valence band of both QD's. A possible
admixing of higher orbital levels caused by the Coulomb
interaction is determined by the parameter $r_{A,B}/a_X$, where
$a_X \simeq 5.4\, {\rm nm}$ is the exciton radius for
CdSe.~\cite{fu:99} For the small QD's in
Ref.~\onlinecite{ouyang:03}, the Coulomb interaction is small
compared to the single-particle level spacing, such that the
admixing of higher orbital levels to the ground state is small as
well. (For details on experimental parameters, see
Sec.~\ref{sec:exp}.) This allows us to describe the coupled QD's
by the Hamiltonian
\begin{equation}
\hat{H}=\hat{H}_0 + \hat{H}_{\rm Coul} + \hat{H}_{T},
\label{eq:ham}
\end{equation}
where
\begin{equation}
\hat{H}_0 = \sum_{\nu = A,B; \sigma = \pm} \left( E_c^\nu
\hat{c}_{c,\sigma}^{\nu\dagger}  \hat{c}_{c,\sigma}^{\nu} +
E_{v}^\nu \hat{c}_{v,\sigma}^{\nu\dagger} \hat{c}_{v,\sigma}^{\nu}
\right) \label{eq:ham-a}
\end{equation}
contains the single-particle levels of uncoupled QD's $\nu=A,B$.
The operators $\hat{c}_c^{\nu,\sigma}$ and
$\hat{c}_v^{\nu,\sigma}$ annihilate an electron in the lowest
level $E_c^\nu$ of the conduction band with spin quantum number
$s_z = \sigma 1 /2$ and the highest level in the valence band,
$E_v^\nu$, with angular momentum $j_z = \sigma 3/2$, respectively,
where $\sigma=\pm$. Here, we have adopted a simple model for the
change in the band structure of CdSe due to the QD confinement. We
assume a spherical QD shape and a splitting of the $j=3/2$ valence
band at the $\Gamma$ point into the heavy hole (hh) and light hole
(lh) subband with total angular momentum projection $j_z=\pm 3/2$
and $j_z=\pm 1/2$, respectively, as obtained, e.g., from the
Luttinger Hamiltonian with an additional anisotropy term for the
crystal field of the hexagonal lattice.~\cite{efros:92} The lh
subband will therefore be neglected in the following. The Coulomb
interaction energy is
\begin{equation}
\hat{H}_{\rm Coul} = \sum_{\nu = A,B} \frac{U_\nu}{2} \left[
\hat{n}_c^\nu (\hat{n}_c^\nu-1) + \hat{n}_v^\nu (\hat{n}_v^\nu-1)
- 2 \hat{n}_c^\nu \hat{n}_v^\nu \right], \label{eq:ham-b}
\end{equation}
where
$\hat{n}^\nu_c=\sum_{\sigma=\pm}\hat{c}_{c,\sigma}^{\nu\dagger}
\hat{c}_{c,\sigma}^{\nu}$ and $\hat{n}^\nu_v = \sum_{\sigma = \pm}
\hat{c}_{v,\sigma}^{\nu}  \hat{c}_{v,\sigma}^{\nu \dagger}$ are
the number operators for electrons in the conduction band level
and holes in the valence band level. $U_\nu \simeq e^2/4 \pi
\epsilon \epsilon_0 r_\nu$ is the characteristic charging energy
of QD $A$ and $B$, respectively. Transfer of spin and charge
between the QD's is accounted for by the transfer Hamiltonian
\begin{equation}
\hat{H}_T = \sum_{\sigma=\pm} \left( t_{c}
\hat{c}_{c,\sigma}^{A\dagger} \hat{c}_{c,\sigma}^{B}
+ t_{v}
\hat{c}_{v,\sigma}^{A\dagger} \hat{c}_{v,\sigma}^{B} + h.c.
\right),
\label{eq:ham-c}
\end{equation}
 where we assume that transfer of electrons through the
$\pi$-conjugated molecule conserves the electron spin both in  the
conduction and the valence band.

The ansatz for the Hamiltonian in
Eqs.~(\ref{eq:ham})--(\ref{eq:ham-c}) is a model in which the
biexciton shift, the exciton fine structure, and the electrostatic
coupling between the QD's have been neglected. We will justify
this in Sec.~\ref{sec:exp} below where we discuss our results for
the experimental parameters of Ref.~\onlinecite{ouyang:03}.
Because the focus of this work is to calculate the FR angle that
results from transfer of electrons between the QD's, we assume for
simplicity that the symmetry axis of the QD's with hexagonal
crystal structure is parallel to the direction of pump and probe
laser pulses. The effect of a random QD orientation  will be
discussed in Sec.~\ref{sec:exp}.

In the following, we analyze the results of
Ref.~\onlinecite{ouyang:03} based on the Hamiltonian
Eq.~(\ref{eq:ham}). This paper is organized as follows. In
Sec.~\ref{sec:trfr}, we calculate the time-resolved FR signal for
an electron wave function which is delocalized over QD's $A$ and
$B$. In Sec.~\ref{sec:exc-exc}, we calculate the FR angle as a
function of probe energy for an initial spin polarization created
by optical pumping. We take into account both electron transfer
processes and the Coulomb interaction and show that these terms
give rise to an exchange splitting of the two-exciton eigenstates.
In Sec.~\ref{sec:doped-exc}, we perform the related analysis for a
system with one spin-polarized excess electron in the QD's. In
Sec.~\ref{sec:exp}, we discuss our results for the parameters of
CdSe QD's coupled by benzene molecules,~\cite{ouyang:03} calculate
the transfer matrix element and spin transfer probabilities. In
Sec.~\ref{sec:concl}, we draw our conclusions.

\section{Time resolved Faraday rotation for coupled Quantum Dots}
\label{sec:trfr}

Before we calculate the FR angle for the general Hamiltonian
Eq.~(\ref{eq:ham}) in Secs.~\ref{sec:exc-exc} and
\ref{sec:doped-exc} below, we first consider time-resolved FR for
a particularly simple case in which a single electron is in a
coherent superposition of states in QD's $A$ and $B$ at time
$t=0$, $|\psi(0) \rangle = (\hat{c}_{c,+}^{B \dagger}+ \alpha
\hat{c}_{c,+}^{A \dagger})| 0 \rangle/ \sqrt{1+\alpha^2}$. We
further assume $t_{c,v}=0$ and $E_c^A=E_c^B$ in Eq.~(\ref{eq:ham})
for $t>0$. Here, $| 0 \rangle$ denotes the vacuum state in which
the valence band in both QD's is filled and the conduction band
states are empty. This simple scenario, although unrealistic
because transfer matrix elements are assumed to vanish after the
initial state $|\psi_0\rangle$ has been prepared, will allow us to
derive simple analytical expressions for the FR angle even in
presence of a magnetic field. The simplifying assumptions
$t_{c,v}=0$ and $E_c^A=E_c^B$ will be lifted in the microscopic
discussion in Secs.~\ref{sec:exc-exc} and \ref{sec:doped-exc}.

The different radii $r_A$ and $r_B$ of the CdSe QD's lead to
different $g$-factors and different Larmor precession frequencies
$\omega_{\nu}= g_\nu \mu_B B_{ext}/
\hbar$,~\cite{gupta:02,rodina:03,schrier:03} where $B_{ext}$ is an
external magnetic field perpendicular to the spin quantization
axis which is given by the symmetry axis of the CdSe QD's, and
$g_\nu$ are the electron $g$-factors for $\nu=A,B$. At time $t$,
\begin{eqnarray}
&& |\psi(t)\rangle = \frac{1}{ \sqrt{1+\alpha^2}} \left[ \cos
(\omega_B t/2)\hat{c}_{c,+}^{B\dagger}
- i \sin (\omega_B t/2)\hat{c}_{c,-}^{B\dagger}  \right. \nonumber \\
&& \hspace*{0.5cm} \left. + \alpha \cos (\omega_A
t/2)\hat{c}_{c,+}^{A\dagger} - i \alpha \sin (\omega_A
t/2)\hat{c}_{c,-}^{A\dagger} \right]| 0 \rangle .
\label{eq:t-state1}
\end{eqnarray}

This time evolution of the electron spin can be detected by FR
because the FR angle $\theta_F$ is determined by the population
imbalance between the $s_z=\pm 1/2$ conduction band states in this
situation.~\cite{hugonnard:94,linder:98,sham:99} For probe pulse
frequency $E/h$, $\theta_F$ is proportional to the difference of
the real parts of the dielectric response functions $\epsilon (E)$
for $\sigma^\pm$ circularly polarized light.~\cite{hugonnard:94}
With the spectral representation of the response functions,
$\theta_F(E)$ is expressed in terms of the transition matrix
elements between the state $|\psi(t)\rangle$ with energy $E_0$ and
all intermediate states $|\psi_i\rangle$ which are virtually
excited by the probe pulse,
\begin{eqnarray}
&& \theta_F(E,t) = C E \sum_{|\psi_i\rangle} \frac{E-(E_i-E_0)}{
[E-(E_i-E_0)]^2+\Gamma^2}  \label{eq:frot-g}
\\ && \hspace*{1cm} \times \left( \left| \langle \psi_i  |
\hat{P}_+|\psi(t) \rangle \right|^2 - \left| \langle \psi_i  |
\hat{P}_-|\psi(t) \rangle \right|^2 \right). \nonumber
\end{eqnarray}
The polarization operators $\hat{P}_\pm=d_A
\hat{c}_{c,\pm}^{A\dagger} \hat{c}_{v,\pm}^A + d_B
\hat{c}_{c,\pm}^{B\dagger} \hat{c}_{v,\pm}^B$ couple to the
$\sigma^\mp$ circularly polarized components of the probe pulse.
$d_\nu$ are the dipole transition matrix elements for transition
from the $j_z = \pm 3/2$ valence band states to the $s_z = \pm
1/2$ conduction band states in  QD's $A$ and $B$. $E_0 = E_c^B$
and $E_i$ are the energy eigenvalues of the initial state and the
intermediate state $|\psi_i\rangle$, respectively, and the level
broadening $\Gamma$ accounts for a finite lifetime of the orbital
levels. The prefactor $C \propto L/(h c n_0)$ is determined by the
size $L$ of the sample and the refraction index $n_0$ of bulk
CdSe.

Because we have assumed an initial state $|\psi(0)\rangle$ with
one electron, all intermediate states $|\psi_i\rangle$ in
Eq.~(\ref{eq:frot-g}) are energy eigenstates with two electrons
and one hole. For $t_{c,v}=0$ in Eq.~(\ref{eq:ham}), these are of
the form $|\psi_i\rangle = \hat{c}_{c,\sigma}^{\nu\dagger}
\hat{c}_{v,\sigma}^\nu \hat{c}_{c,\sigma^\prime}^{\nu^\prime
\dagger}|0\rangle$ with $\sigma,\sigma^\prime=\pm$ and
$\nu,\nu^\prime=A,B$. Pauli blocking prohibits the creation of an
exciton with electron spin $\sigma 1/2$ if the conduction band
level is already occupied by an electron with the same spin. The
resulting difference in transition matrix elements for $
\hat{P}_+$ and $\hat{P}_-$ is proportional to the population
imbalance of the $s_z = \pm 1/2$ levels. For a probe pulse at time
$t$, from Eq.~(\ref{eq:frot-g}) we obtain directly
\begin{eqnarray}
\theta_F (E,t) &=& \frac{CE}{1+\alpha^2} \left[ d_B^2
\frac{E-E_{X}^B}{(E-E_{X}^B)^2+\Gamma^2}\cos (\omega_B t) \right.
\nonumber \\  && \left. + \alpha^2 d_A^2
\frac{E-E_{X}^A}{(E-E_{X}^A)^2+\Gamma^2} \cos (\omega_A t)\right],
\label{eq:frot1a}
\end{eqnarray}
where $E_{X}^\nu=E_{c}^\nu - E_v^\nu - U_\nu$ is the exciton
energy for QD $\nu$.

$\theta_F(E,t)$ shows coherent oscillations
with frequencies $\omega_A$ and $\omega_B$ caused by the electron
spin precessing around the external magnetic field. In reality,
these coherent oscillations are exponentially damped with a spin
dephasing rate $\Gamma_S$ which is typically much smaller than the
orbital dephasing rate, $\Gamma_S \ll \Gamma$. Taking into account
spin dephasing, the Fourier transform of the time-resolved FR
signal as a function of the probe pulse energy $E$ and the Fourier
frequency $\omega$ is
\begin{eqnarray}
&& \theta_F (E,\omega) = \frac{CE}{1+\alpha^2} \nonumber \\
&& \hspace*{0.5cm} \times  \left[ d_B^2
\frac{E-E_{X}^B}{\left(E-E_{X}^B\right)^2+\Gamma^2}
\frac{\Gamma_S}{\left(\omega-\omega_B \right)^2+\Gamma_S^2}
\right. \nonumber
\\  && \left.
 \hspace*{0.5cm} + \alpha^2 d_A^2 \frac{E-E_{X}^A}{(E-E_{X}^A)^2+\Gamma^2}
\frac{\Gamma_S}{\left(\omega-\omega_A \right)^2+\Gamma_S^2}
\right]. \label{eq:frot1b}
\end{eqnarray}

\begin{figure}
\centerline{\mbox{\includegraphics[width=7.3cm]{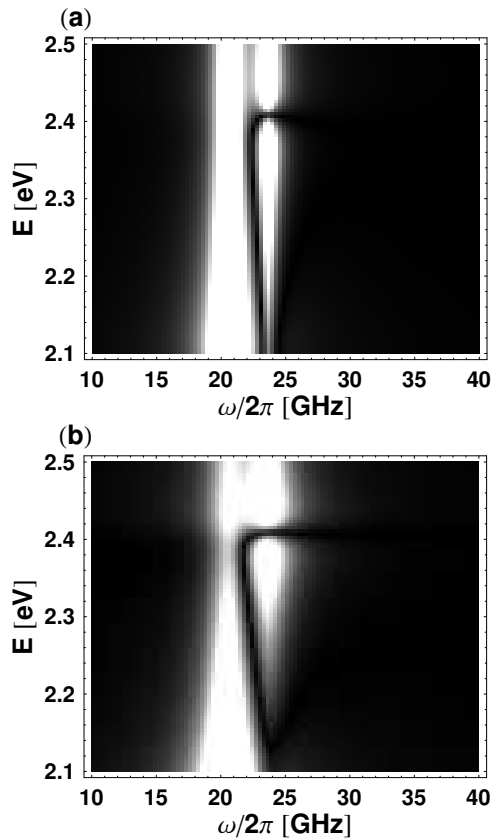}}}
\caption{(a) Grayscale plot of the FR angle $|\theta_F|$ [given in
Eq.~(\ref{eq:frot1b})] in arbitrary units as function of probe
pulse frequency $E/h$ and frequency $f=\omega/2\pi$. We have
chosen the parameters~\cite{ouyang:03} $E_X^A=2.41 \,{\rm eV}$,
$E_X^B=2.06 \,{\rm eV}$, $\omega_A/2\pi = 23.6 \, {\rm GHz}$,
$\omega_B / 2\pi = 20.6 \, {\rm GHz}$, $\Gamma = 0.05 \, {\rm
eV}$, $\Gamma_S/2 \pi = 0.5 \, {\rm GHz}$, and $\alpha^2= 0.2$.
(b) For $\Gamma = 0.035 \, {\rm eV}$, $\Gamma_S/2 \pi = 1.2 \,
{\rm GHz}$, and $\alpha^2= 0.4$, pronounced features caused by the
interplay of the two terms in Eq.~(\ref{eq:frot1b}) become more
clearly visible. In particular, the FR signal at $\omega \simeq
\omega_B$ vanishes and reappears as a function of probe pulse
frequency $E$.
 } \label{Fig1}
\end{figure}

$\theta_F(E,\omega)$ shows characteristic features for $E\simeq
E_X^\nu$ and $\omega \simeq \omega_\nu$. The two terms in
Eq.~(\ref{eq:frot1b}) describe the dielectric response due to
virtual creation of an exciton in QD $A$ and $B$, respectively.
For $E_X^B \leq E \leq E_X^A$, they have different sign and may
cancel. Figure~\ref{Fig1}(a) shows a grayscale plot of $|\theta_F
(E,\omega) |$ for the experimental values $E_X^B = 2.06 \,{\rm
eV}$, $E_X^A = 2.41 \, {\rm eV}$, $\Gamma = 0.05 \, {\rm eV}$, and
$\Gamma_S/2 \pi = 0.5 \, {\rm GHz}$, assuming $d_A^2/d_B^2=1$ and
$\alpha^2 = 0.2$. For Fig.~\ref{Fig1}(b), $\Gamma = 0.035 \, {\rm
eV}$, and $\Gamma_S/2 \pi = 1.2 \, {\rm GHz}$, and $\alpha^2 =
0.4$. One of the most characteristic features of the experimental
data (Fig. 2C in Ref.~\onlinecite{ouyang:03}) is that $|\theta_F
(E,\omega) |$ vanishes and reappears as a function of probe pulse
frequency $E$ for $\omega \simeq \omega_B$. This feature is also
present in the theoretical result and can be traced back to the
superposition of two response functions in Eq.~(\ref{eq:frot1b}).
More specifically, for $\omega \simeq \omega_B$ and $E\simeq
E_X^A-\Gamma$, the two terms in Eq.~(\ref{eq:frot1b}) have
opposite sign and cancel for sufficiently large $\alpha$.

Above, we have assumed that the electron delocalized over both
QD's at $t=0$ retains spatial coherence. For rapid decoherence of
the orbital part of the wave function, the initial state is
described by the density matrix $\hat{\rho} =
\left(\hat{c}_{c,+}^{B\dagger}|0\rangle \langle
0|\hat{c}_{c,+}^{B}+ \alpha^2 \hat{c}_{c,+}^{A\dagger}|0\rangle
\langle 0|\hat{c}_{c,+}^{A} \right)/(1+\alpha^2)$. The FR signal
in this case is the incoherent superposition of the FR signals for
QD $A$ and $B$, and is identical to the results in
Eqs.~(\ref{eq:frot1a}) and (\ref{eq:frot1b}). Hence, a FR signal
as shown in Fig.~\ref{Fig1} does not allow one to distinguish
coherent from incoherent spatial superpositions.

\section{Optical spin injection}
\label{sec:exc-exc}

In the preceding section, $\theta_F(E)$ was calculated for the
simple case of a single electron delocalized over the coupled
QD's. So far, we have also neglected that all intermediate states
$|\psi_i\rangle$ in Eq.~(\ref{eq:frot-g}) that are virtually
excited by the probe pulse will be modified by finite transfer
energies $t_{c,v}$. We next turn to a microscopic analysis in
which we take into account $t_{c,v} \neq 0$ also for the
intermediate states.

In Ref.~\onlinecite{ouyang:03}, the initial state prepared by
optical pumping is a one-exciton state. Similar to the analysis in
Sec.~\ref{sec:trfr} above, the FR angle as a function of probe
energy is proportional to the difference of the dielectric
response functions for $\sigma^\pm$ circularly polarized light
[Eq.~(\ref{eq:frot-g})]. In order to evaluate this expression,
both the initial one-exciton state and all intermediate
two-exciton states which are virtually excited by the probe pulse
must be calculated for the coupled QD's. In this Section, we first
calculate the one-exciton energy eigenstate of the coupled QD's
prepared by the pump pulse and subsequently identify all
two-exciton eigenstates $|\psi_i\rangle$ which are virtually
excited by the probe pulse. Our analysis is based on perturbation
theory in the transfer energies and is valid if $ |t_{c,v}|$ is
the smallest energy scale, $ |t_{c,v}| \ll \delta E_c, |\delta E_v
|, U_A, U_B, |\delta E_{c,v} \pm U_{A,B} |$. Here, we have defined
the energy differences $\delta E_c = E_c^A - E_c^B \geq 0$ and
$\delta E_v = E_v^A - E_v^B \leq 0$ between the conduction and
valence band levels of QD's $A$ and $B$.

In Ref.~\onlinecite{ouyang:03}, an initial
spin polarization was created by optical pumping.
For $t_{c,v}=0$, the states $\hat{c}_{c,\sigma}^{\nu \dagger}
\hat{c}_{v,\sigma}^\nu |0 \rangle$ are one-exciton eigenstates with energy
eigenvalues
\begin{equation}
E_X^{\nu(0)}=E_{c}^\nu-E_v^\nu-U_\nu  \label{eq:1exc-unp}
\end{equation}
which are prepared by absorption of a $-\sigma$ circularly
polarized pump pulse. To first order in the transfer energies
$t_{c,v}$, the energy eigenstates are
\begin{subequations}
\label{eq:1-exc-st}
\begin{eqnarray}
|X_{A,\sigma} \rangle &=& \hat{c}_{c,\sigma}^{A\dagger}
 \hat{c}_{v,\sigma}^{A} |0\rangle
+ \left(\frac{t_c}{\delta E_c-U_A}\hat{c}_{c,\sigma}^{B\dagger}
 \hat{c}_{v,\sigma}^{A} \right. \nonumber \\ && \left. \hspace*{1cm} +
\frac{t_v}{\delta E_v +U_A}\hat{c}_{c,\sigma}^{A\dagger}
 \hat{c}_{v,\sigma}^{B} \right) |0\rangle,  \label{eq:1-exc-sta} \\
|X_{B,\sigma} \rangle &=& \hat{c}_{c,\sigma}^{B\dagger}
 \hat{c}_{v,\sigma}^{B} |0\rangle
+ \left(-\frac{t_c}{ \delta E_c +
U_B}\hat{c}_{c,\sigma}^{A\dagger}
 \hat{c}_{v,\sigma}^{B} \nonumber \right. \\ && \left. \hspace*{1cm} -
\frac{t_v}{\delta E_v-U_B}\hat{c}_{c,\sigma}^{B\dagger}
 \hat{c}_{v,\sigma}^{A} \right) |0\rangle , \label{eq:1-exc-stb}
\end{eqnarray}
\end{subequations}
with eigenenergies
\begin{subequations}
\label{eq:1-exc-en}
\begin{eqnarray}
E_X^A &=& E_X^{A(0)}+ \frac{t_c^2}{\delta E_c-U_A}
- \frac{t_v^2}{\delta E_v+U_A}, \label{eq:1-exc-ena} \\
E_X^B &=& E_X^{B(0)}- \frac{t_c^2}{\delta E_c +U_B} +
\frac{t_v^2}{\delta E_v-U_B}. \label{eq:1-exc-enb}
\end{eqnarray}
\end{subequations}

As expected, the eigenenergies are shifted due to the
delocalization of electrons and holes over the coupled QD's. The
exciton states in Eq.~(\ref{eq:1-exc-st}) are the only one-exciton
states which can be prepared by the absorption of a photon with
circular polarization $-\sigma$ if the photon is incident along
the hexagonal axis of the CdSe crystal structure. However, a
photon with energy $E \simeq E_X^B$ no longer creates an exciton
only in QD B, but an exciton in which electron and hole are
delocalized over the coupled QD system. This delocalization of the
quantum mechanical wave function is consistent with the short time
scale for spin transfer observed experimentally.~\cite{ouyang:03}

We now turn to the calculation of the FR angle, assuming that the
pump pulse has prepared an initial state $|\psi\rangle =
|X_{B,+}\rangle$. The evaluation of the dielectric response
function will require us to calculate all two-exciton states that
are virtually excited by the probe pulse. Interesting features in
the FR signal effected by spin transfer are of order $t_{c,v}^2$.
In order to keep the following expressions simple, we assume that
spin is transferred between the conduction band states and set
$t_v = 0$. Then, only the seven states $|A_+ B_+ \rangle$,
$|T_0\rangle$, $|S\rangle$, $|B_+ B_- \rangle$,
$|\tilde{T}_0\rangle$, $|\tilde{S}\rangle$, and $|\widetilde{B_+
B_-} \rangle$ listed below and in Appendix~\ref{ap:2exc} have
finite matrix elements up to ${\cal O}(t_c^2)$ with $\hat{P}_\pm
|X_{B,+}\rangle$. For $\delta E_v + U_A \neq 0$, only the
eigenenergies of $|A_+ B_+ \rangle$, $|T_0\rangle$, and
$|S\rangle$ are close to the excitation energy of a probe pulse
with frequency $E/h \simeq E_X^A/h$. Hence, these states dominate
the spectral representation in Eq.~(\ref{eq:frot-g}).~\cite{rem2}

The polarization operator $\hat{P}_+$ induces transitions from the
initial state $|X_{B,+}\rangle$ to
\begin{equation} |A_+ B_+ \rangle = \hat{c}_{c,+}^{A\dagger}
\hat{c}_{c,+}^{B\dagger} \hat{c}_{v,+}^{A}  \hat{c}_{v,+}^{B}|0
\rangle \label{eq:2exc-1s}
\end{equation}
with energy eigenvalue
\begin{equation}
E_{A_+ B_+ } = E_X^{A(0)} +  E_X^{B(0)}. \label{eq:2exc-1e}
\end{equation}
The notation indicates that two electrons with the same spin $s_z
= 1/2$ occupy the conduction band states in  QD's A and B,
respectively, and form a spin triplet state. $|A_+ B_+ \rangle$ is
an exact eigenstate of the Hamiltonian even for $t_c \neq 0$
because transfer of the conduction band electrons is blocked by
Pauli's exclusion principle. The matrix element $\langle A_+ B_+ |
\hat{P}_+ |X_{B_+} \rangle$ is the only finite matrix element of
the operator $\hat{P}_+ $.

Finite matrix elements for $\hat{P}_-$ come from the states in
which the electrons in the conduction band level form a spin
triplet and singlet, respectively,
\begin{subequations}
\label{eq:2exc-2s}
\begin{eqnarray}
|T_0 \rangle &=& \frac{1}{\sqrt{2}} \left(\hat{c}_{c,-}^{A\dagger}
\hat{c}_{c,+}^{B\dagger} + \hat{c}_{c,+}^{A\dagger}
\hat{c}_{c,-}^{B\dagger} \right) \hat{c}_{v,-}^{A}
\hat{c}_{v,+}^{B}
|0 \rangle, \label{eq:2exc-2sa}  \\
|S\rangle &\propto & \frac{1}{\sqrt{2}}
\left(\hat{c}_{c,-}^{A\dagger} \hat{c}_{c,+}^{B\dagger} -
\hat{c}_{c,+}^{A\dagger} \hat{c}_{c,-}^{B\dagger} \right)
\hat{c}_{v,-}^{A}  \hat{c}_{v,+}^{B}
|0 \rangle \label{eq:2exc-2sb} \\
&& + \sqrt{2} \left(\frac{t_c}{\delta E_c + U_B} \hat{c}_{c,+}^{A\dagger}
\hat{c}_{c,-}^{A\dagger}  - \frac{t_c}{\delta E_c - U_A}
\hat{c}_{c,+}^{B\dagger} \hat{c}_{c,-}^{B\dagger} \right)  \nonumber \\
&& \hspace*{2cm} \times \hat{c}_{v,-}^{A}  \hat{c}_{v,+}^{B}  |0 \rangle,
\nonumber
\end{eqnarray}
\end{subequations}
and the holes with $j_z = -3/2$ and $j_z=+3/2$ are localized in
QD's $A$ and $B$, respectively. Note that the projection of the
total conduction band spin onto the spin quantization axis
vanishes for the triplet state $|T_0\rangle$. The normalization
constant for $|S\rangle$ is defined by $\langle S|S\rangle = 1$.
The eigenenergies
\begin{subequations}
 \label{eq:2exc-2e}
\begin{eqnarray}
E_{T_0 } &=&  E_X^{A(0)} +  E_X^{B(0)}, \label{eq:2exc-2ea} \\
E_{S} &=&  E_X^{A(0)} +  E_X^{B(0)}  \label{eq:2exc-2eb}
\\ && \hspace{0.5cm} + 2 t_c^2 \left( \frac{1}{\delta E_c - U_A}
- \frac{1}{\delta E_c + U_B} \right) \nonumber
\end{eqnarray}
\end{subequations} show an energy offset which is caused by the
inter-dot exchange coupling.~\cite{loss:98,gywat:02} The energies
of $|A_+ B_+ \rangle$ and $|T_0\rangle$ are not shifted by
electron transfer because of Pauli blocking
 and destructive interference of transfer paths,
respectively.

\begin{figure}
\centerline{\mbox{\includegraphics[width=8.3cm]{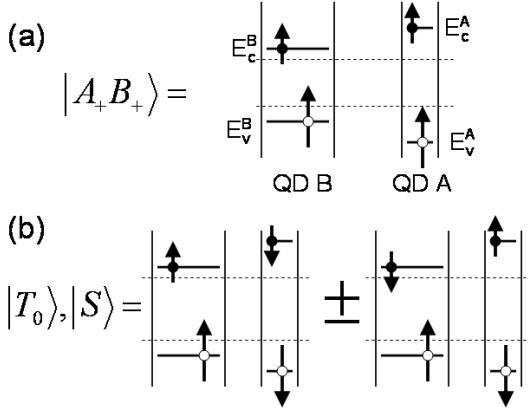}}}
\caption{Schematic representation of the spin configurations (in
the electron picture) for states (a) $|A_+ B_+ \rangle$ and (b)
$|S\rangle$, $|T_0\rangle$ to leading order in $t_c$. The dashed
lines represent the conduction and valence band edge in bulk
CdSe.} \label{Fig2}
\end{figure}

The state
\begin{eqnarray}
&& |B_+ B_- \rangle \propto \left[ \hat{c}_{c,+}^{B\dagger}
\hat{c}_{c,-}^{B\dagger} + \frac{t_c}{\delta E_c - U_A} \left(
\hat{c}_{c,-}^{A\dagger} \hat{c}_{c,+}^{B\dagger} -
\hat{c}_{c,+}^{A\dagger} \hat{c}_{c,-}^{B\dagger} \right) \right.
\nonumber \\ && \left. \hspace*{0.5cm} + \frac{2
t_c^2\hat{c}_{c,+}^{A\dagger} \hat{c}_{c,-}^{A\dagger} }{ (\delta
E_c - U_A)(2 \delta E_c - U_A + U_B)} \right] \hat{c}_{v,-}^{A}
\hat{c}_{v,+}^{B}  |0 \rangle  \label{eq:2exc-3s}
\end{eqnarray}
with
\begin{equation}
E_{B_+ B_- } = E_X^{B(0)} + E_c^B - E_v^A  - 2 \frac{t_c^2}{\delta
E_c - U_A} \label{eq:2exc-3e}
\end{equation}
is offset in energy from $E_X^{A(0)} +  E_X^{B(0)}$ even to zeroth
order in $t_c$ and does not contribute significantly to
$\theta_F(E)$ for $E \simeq E_X^A$. The three states in
Eqs.~(\ref{eq:2exc-2s}) and (\ref{eq:2exc-3s}) provide the
dominant terms in the spectral representation for $\theta_F$ in
Eq.~(\ref{eq:frot-g}). In particular, they exhaust the sum rule
$\sum_{ |\psi_i \rangle} | \langle \psi_i |
\hat{c}_{c,-}^{A\dagger} \hat{c}_{v,-}^{A} | X_{B,+} \rangle |^2 =
1$ up to ${\cal O}(t_c^2)$. In Fig.~\ref{Fig2}, the spin
configurations for $|A_+B_+\rangle$, $|S\rangle$, and
$|T_0\rangle$ are shown schematically.

From Eqs.~(\ref{eq:1-exc-stb})--(\ref{eq:2exc-3e}), the FR angle
$\theta_F$ is readily evaluated. We denote the electron transfer
probability from QD $\nu$ to QD $\nu^\prime$ by $p_{\nu
\rightarrow \nu^\prime}$. We obtain
\begin{subequations}
\label{eq:tun-prob}
\begin{eqnarray}
p_{A\rightarrow B} & = &  \left(\frac{t_c}{\delta E_c - U_A} \right)^2 ,
\label{eq:tun-proba} \\
p_{B\rightarrow A} & = &  \left(\frac{t_c}{\delta E_c + U_B}
\right)^2. \label{eq:tun-probb}
\end{eqnarray}
\end{subequations}
For the transition matrix elements of the dipole operators in
Eq.~(\ref{eq:frot-g}), we obtain in terms of the transfer
probabilities
\begin{subequations}
\label{eq:2exc-matr}
\begin{eqnarray}
&& |\langle A_+ B_+|\hat{P}_+ | X_{B,+} \rangle|^2 =\left( 1-
p_{B\rightarrow A} \right) d_A^2, \label{eq:2exc-matr-a} \\
&& |\langle T_0|\hat{P}_- | X_{B,+} \rangle|^2 = \frac{1-
p_{B\rightarrow A}}{2}  d_A^2, \label{eq:2exc-matr-b} \\
&& |\langle S|\hat{P}_- | X_{B,+} \rangle|^2 = \frac{1+
p_{B\rightarrow A} - 2 p_{A\rightarrow B}}{2} d_A^2,
\label{eq:2exc-matr-c} \\
&& |\langle B_+B_-|\hat{P}_- | X_{B,+} \rangle|^2 =
p_{A\rightarrow B}
  d_A^2. \label{eq:2exc-matr-d}
\end{eqnarray}
\end{subequations}

Because of the exchange splitting $E_{T_0 } -E_{S}$ between
conduction band triplet and singlet states, finite transfer
probabilities $p_{A\rightarrow B}$ and $p_{B\rightarrow A}$ lead
to pronounced features in the FR angle as a function of the probe
pulse frequency $E/h$. For probe energies $E_{T_0 B} = E_{T_0 } -
E_X^B \leq E \leq E_{SB} = E_{S} - E_X^B$, the FR signal varies
strongly with energy and is given by
\begin{eqnarray}
&& \theta_F (E) = \frac{C E d_A^2}{2} \left[
\left(1- p_{B \rightarrow A} \right)
 \frac{E-E_{T_0 B}}{(E-E_{T_0 B})^2 + \Gamma^2}
\right. \hspace*{0.5cm} \nonumber
\\ && \hspace*{0.2cm} \left. -   \left(
1+ p_{B \rightarrow A} - 2 p_{A \rightarrow B}  \right)
 \frac{E-E_{SB}}{(E-E_{SB})^2 + \Gamma^2}  \right].
\label{eq:frot-2}
\end{eqnarray}
For $|E- E_{SB}| \gtrsim |E_{T_0 } -E_{S}|$, Eq.~(\ref{eq:frot-2})
simplifies to
\begin{equation}
\theta_F (E) \simeq C E d_A^2
\frac{E-E_X^{A(0)}}{\left(E-E_X^{A(0)}\right)^2 + \Gamma^2}
\left(p_{A \rightarrow B} - p_{B\rightarrow A}\right).
\label{eq:frot-3}
\end{equation}
This result is surprising because the FR angle is not only
determined by the probability $p_{B\rightarrow A}$ that the
electron created by the pump pulse has been transferred to QD $A$.
Rather, even the \emph{sign} of the FR angle depends on the
parameters $\delta E_{c}$ (and $\delta E_{v}$ if transfer between
valence band states is included) and $U_{A,B}$. $\theta_F \geq 0$
for $|\delta E_c -U_A| \geq |\delta E_c +U_B| $, and $\theta_F
\leq 0$ for $|\delta E_c -U_A| \leq |\delta E_c +U_B|$. Although
counterintuitive at first sight, this can be readily understood
from the one- and two-exciton eigenstates. The matrix element for
the virtual creation of an exciton with $s_z = 1/2$, $j_z=3/2$ in
QD $A$ is reduced by the probability $p_{B\rightarrow A}$ that the
conduction band electron created by the pump pulse in $B$ has been
transferred to $A$. In this case, it blocks the creation of a
second exciton with the same spin. The transition matrix element
for the creation of an exciton with $s_z = -1/2$, $j_z=- 3/2$  is
reduced by the probability $p_{A \rightarrow B}$ that the electron
with spin $s_z=-1/2$ in the conduction band state of QD $A$ is
transferred to QD $B$. This transfer process is not prohibited by
Pauli blocking and leads to the virtual occupation of $|B_+ B_-
\rangle$ which is energetically far off resonance. The interplay
of both processes results in Eq.~(\ref{eq:frot-3}).

\begin{figure}
\centerline{\mbox{\includegraphics[width=7.0cm]{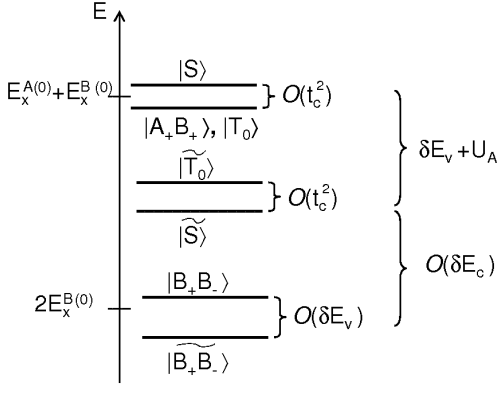}}}
\caption{Energy level scheme of all two-exciton eigenstates
discussed in the text. The eigenenergies fall into three groups
which are split by terms of order ${\cal O}(t_c^2)$ or ${\cal
O}(\delta E_v)$. For the QD's used in Ref.~\onlinecite{ouyang:03},
$\delta E_v + U_A \simeq 0$, and the five states $|A_+B_+\rangle$,
$|T_0\rangle$, $|S\rangle$, $|\tilde{T}_0\rangle$, and
$|\tilde{S}\rangle$ are nearly degenerate.} \label{Fig3}
\end{figure}

\begin{figure}
\centerline{\mbox{\includegraphics[width=8.3cm]{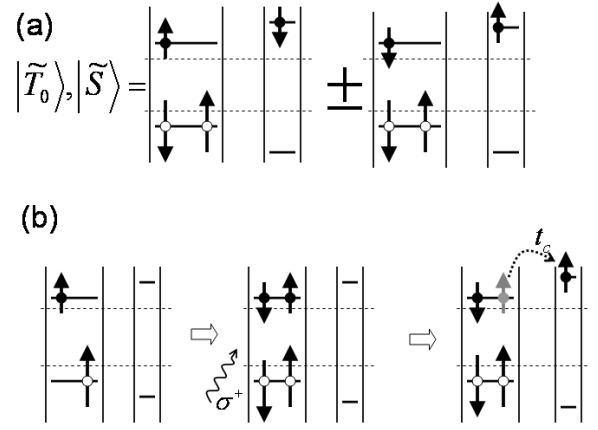}}}
\caption{(a) Schematic representation of the spin configurations
for the states $|\tilde{S}\rangle$, $|\tilde{T}_0\rangle$ to
leading order in $t_c$. (b) Transitions between an initial state
$| X_{B,+} \rangle$ and $|\tilde{S}\rangle$, $|\tilde{T}_0\rangle$
are effected by the absorption of a $\sigma^+$ polarized probe
photon and subsequent tunneling of one conduction band electron.}
\label{Fig4}
\end{figure}

Our derivation of Eq.~(\ref{eq:frot-3})  was based on the
assumption that $t_c$ is the smallest energy scale in the system.
As will be discussed in Sec.~\ref{sec:exp} below, for the
experimental parameters in Ref.~\onlinecite{ouyang:03}, $\delta
E_v + U_A \simeq 0$. For $t_v=0$, this does not lead to
divergencies in the perturbative expansion in $t_c$. However,
these special parameters require that two additional two-exciton
states are taken into account for the calculation of $\theta_F
(E)$ because they are nearly degenerate with $|A_+B_+\rangle$,
$|S\rangle$, and $|T_0\rangle$ (see Fig.~\ref{Fig3}): The states
$|\tilde{S}\rangle$ and $|\tilde{T}_0\rangle$ defined in
Eq.~(\ref{eq:2exc-4s}) have finite overlap matrix elements with
$\hat{P}_- |X_{B,+}\rangle$,
\begin{subequations}
\label{eq:2exc-matrb}
\begin{eqnarray}
|\langle \tilde{T}_0|\hat{P}_- | X_{B,+} \rangle|^2 & = & \frac{
p_{B\rightarrow A} d_B^2}{2} , \label{eq:2exc-matrb-a} \\
|\langle \tilde{S}|\hat{P}_- | X_{B,+} \rangle|^2 & = & \frac{
p_{B\rightarrow A} d_B^2}{2}. \label{eq:2exc-matrb-b}
\end{eqnarray}
\end{subequations}
The spin configurations for the states $|\tilde{S}\rangle$ and
$|\tilde{T}_0\rangle$ are shown schematically in
Fig.~\ref{Fig4}(a). Note that both holes occupy the valence band
states of QD $B$. The accidental degeneracy of $|\tilde{S}\rangle$
and $|\tilde{T}_0\rangle$ with $|S\rangle$ and $|T_0\rangle$
arises because, for the parameters of Ref.~\onlinecite{ouyang:03},
the decrease in orbital energy $\delta E_v$ is comparable to the
increase in Coulomb energy $U_A$. Transitions between an initial
state $| X_{B,+} \rangle$ and $|\tilde{S}\rangle$,
$|\tilde{T}_0\rangle$, are two-step processes. A $\sigma^+$
polarized probe photon creates an exciton with $s_z=-1/2$ and $j_z
= -3/2$ in $B$, and one of the conduction band electrons in $B$ is
subsequently transferred to $A$. These processes are shown
schematically in Fig.~\ref{Fig4}(b).

Taking into account all two-exciton states with energies
$\left|E_i - \left(E_{X}^{A(0)} + E_{X}^{B(0)} \right)\right |
\lesssim \max[|\delta E_v+U_A|, |E_{T_0} -E_{S}|]$, the FR angle
is
\begin{eqnarray}
&& \theta_F (E) = \frac{CE}{2} \left\{ d_A^2  \left[
\left(1- p_{B \rightarrow A} \right)
 \frac{E-E_{T_0 B}}{(E-E_{T_0 B})^2 + \Gamma^2}
\right. \right. \nonumber
\\ && \left. \hspace*{0.5cm} -   \left(
1+ p_{B \rightarrow A} - 2 p_{A \rightarrow B}  \right)
 \frac{E-E_{SB}}{(E-E_{SB})^2 + \Gamma^2}  \right] \label{eq:frot-4} \\
&&  \left. - d_B^2 p_{B \rightarrow A} \left[
 \frac{E-E_{\tilde{T}_0 B}}{(E-E_{\tilde{T}_0 B})^2 + \Gamma^2} +
\frac{E-E_{\tilde{S}B}}{(E-E_{\tilde{S}B})^2 + \Gamma^2}
\right] \right\}. \nonumber
\end{eqnarray}
The energy differences $E_{\tilde{T}_0 B}=E_{\tilde{T}_0 }-E_X^B$
and $E_{\tilde{S}B}=E_{\tilde{S} }-E_X^B$ are given by the
eigenenergies in Eq.~(\ref{eq:2exc-4e}). For $ |E-E_{T_0 B} |,
|E-E_{SB} | \ll |E-E_{\tilde{T}_0 B} |, |E-E_{\tilde{S}B} |$,
Eq.~(\ref{eq:frot-4}) simplifies to
Eq.~(\ref{eq:frot-2}).~\cite{rem3}

Above, we have only considered $t_c \neq 0$ and $t_v = 0$, i.e., a
scenario in which electrons in the valence band remain localized
in the QD's while electrons in conduction band states can be
transferred. The case $t_v \neq 0$ and $t_c = 0$ can be mapped
onto the problem discussed above by mapping electrons onto holes,
i.e., by interchanging $c$ and $v$ in above expressions. In
particular, Eqs.~(\ref{eq:frot-2}) and (\ref{eq:frot-3}) remain
valid if the transfer probabilities for electrons are replaced by
the corresponding values for holes, e.g., $p_{A\rightarrow B}=
[t_v/(\delta E_v + U_A)]^2$, and the energy eigenvalues are
calculated for transfer in the valence rather than the conduction
band.

In the limit of small QD's with similar sizes, $U_{A,B} \gg
t_{c,v} \gg \delta E_c, |\delta E_v|$, configurations in which
electrons and holes occupy different QD's are strongly suppressed.
If $t_{c,v}/U_{A,B} \simeq 0$ but $t_c
t_v/U_{A,B}\left(E_X^{A(0)}- E_X^{B(0)}\right)$ remains finite, a
joint transfer of electron and hole via a virtual intermediate
state is possible. Evidence for this coherent delocalization of an
exciton has been reported for QD's of similar
sizes.~\cite{schedelbeck:97,bayer:01} In contrast, tunneling of
excitons between a pair of QD's with different sizes is incoherent
if the orbital decoherence rate is comparable to the exciton
tunneling rate.~\cite{heitz:98,takeuchi:00,robinson:01,seufert:01}

\section{Doping of coupled Quantum Dots}
\label{sec:doped-exc}

In the last Section, we have analyzed the FR angle for an initial
spin population created by optical pumping, the method used in
Ref.~\onlinecite{ouyang:03}. We now calculate the FR angle
$\theta_F(E)$ for the case that the initial spin density is
carried by an excess electron rather than an exciton. Spin
injection could be achieved, e.g., by doping one CdSe QD with a
single donor atom. For a chemical potential $E_c^B \leq \mu \leq
E_c^A, E_c^B + U_B$, the conduction band level of QD $B$ is filled
with one electron while QD $A$ remains empty. The excess electron
can be spin polarized by cooling in presence of a magnetic field.
Again, we set $t_v = 0$ to keep our results transparent.

The transfer matrix element for the conduction band level
leads to the delocalization
of the excess electron in QD $B$,
\begin{equation}
|e_{B,\sigma} \rangle  = \left[ 1+\left(\frac{t_c}{\delta E_c} \right)^2
\right]^{-1/2}
\left(\hat{c}^{B\dagger}_{c,\sigma}
- \frac{t_c}{\delta E_c} \hat{c}^{A\dagger}_{c,\sigma} \right) |0 \rangle
\label{eq:1-el-sb}
\end{equation}
with eigenenergy $E^B = E_c^B - t_c^2/\delta E_c$. Note that the
energy shift is different from the one found for the exciton
because there is no Coulomb attraction between electron and hole
in the present case.

We calculate the FR angle for an initial state $|e_{B,+} \rangle $
and probe energy $E \simeq E_X^A$. Similar to the analysis in
Sec.~\ref{sec:exc-exc}, three intermediate states dominate the
spectral representation for $\theta_F(E)$. These states are the
following.
\begin{equation}
|A_+B_+^{-} \rangle = \hat{c}_{c,+}^{A\dagger}
\hat{c}_{c,+}^{B\dagger} \hat{c}_{v,+}^{A}  |0 \rangle
\label{eq:eexc-1s}
\end{equation}
with energy eigenvalue
\begin{equation}
E_{A_+B_+^{-} } = E_X^{A(0)} +  E^B_c \label{eq:eexc-1e}
\end{equation}
is populated by creation of an exciton with conduction and valence
band spins $s_z=1/2$ and $j_z = 3/2$, respectively.~\cite{rem4}
Virtual creation of an exciton with $s_z = -1/2$ and $j_z=-3/2$
leads to transitions to the spin triplet and singlet states
\begin{subequations}
\label{eq:eexc-2s}
\begin{eqnarray}
|T_0^{-}\rangle &=& \frac{1}{\sqrt{2}}
\left(\hat{c}_{c,-}^{A\dagger} \hat{c}_{c,+}^{B\dagger} +
\hat{c}_{c,+}^{A\dagger} \hat{c}_{c,-}^{B\dagger} \right)
\hat{c}_{v,-}^{A}
|0 \rangle \label{eq:eexc-2s-a} \\
|S^{-}\rangle &\propto & \frac{1}{\sqrt{2}}
\left(\hat{c}_{c,-}^{A\dagger} \hat{c}_{c,+}^{B\dagger} -
\hat{c}_{c,+}^{A\dagger} \hat{c}_{c,-}^{B\dagger} \right)
\hat{c}_{v,-}^{A}
|0 \rangle \label{eq:eexc-2s-b} \\
&& + \sqrt{2} \left(\frac{t_c}{\delta E_c}
\hat{c}_{c,+}^{A\dagger} \hat{c}_{c,-}^{A\dagger}  -
\frac{t_c}{\delta E_c - U_A - U_B}
\hat{c}_{c,+}^{B\dagger} \hat{c}_{c,-}^{B\dagger} \right)  \nonumber \\
&& \hspace*{2cm} \times \hat{c}_{v,-}^{A}    |0 \rangle,
\nonumber
\end{eqnarray}
\end{subequations}
where the normalization constant for $|S^{-}\rangle$ is determined
by $\langle S^{-}|S^{-}\rangle = 1$. The eigenenergies
\begin{subequations}
 \label{eq:eexc-2e}
\begin{eqnarray}
E_{T_0^{-}} &=&  E_X^{A(0)} +  E_c^{B}, \label{eq:eexc-2e-a} \\
E_{S^{-}} &=&  E_X^{A(0)} +  E_c^{B} \label{eq:eexc-2e-b}
\\ && \hspace*{0.5cm} + 2 t_c^2 \left( \frac{1}{\delta E_c - U_A - U_B} -
\frac{1}{\delta E_c} \right) \nonumber
\end{eqnarray}
\end{subequations}
are split by the exchange coupling of the conduction band levels.
Further, there are several states with energies differing from
$E_X^{A(0)}+E_c^{B}$ (see Appendix~\ref{ap:eexc}). For probe pulse
energies $E \simeq E_X^{A(0)}$ and $|\delta E_v + U_A - U_B|
\gtrsim \Gamma$, $\theta_F(E)$ is dominated by virtual excitations
into the states $|A_+B_+^{-}\rangle$, $|T_0^{-}\rangle$, and
$|S^{-}\rangle$. In this case, all other energy eigenstates with
two conduction band electrons and one hole listed in
Appendix~\ref{ap:eexc} are energetically far off resonance and can
be neglected.

The transition matrix elements of the polarization operators
$\hat{P}_{\pm}$ between $|e_{B,+} \rangle$ and the states
Eqs.~(\ref{eq:eexc-1s}) and (\ref{eq:eexc-2s}) are readily
evaluated. The probabilities for electron transfer between the
QD's are now given by
\begin{subequations}
\label{eq:tun-prob2}
\begin{eqnarray}
p_{B\rightarrow A}^{-} & = &  \left(\frac{t_c}{\delta E_c}
\right)^2 ,
\label{eq:tun-prob2-a} \\
p_{A\rightarrow B}^{-} & = &  \left(\frac{t_c}{\delta E_c - U_A -
U_B} \right)^2. \label{eq:tun-prob2-b}
\end{eqnarray}
\end{subequations}
Then,
\begin{subequations}
\label{eq:eexc-matr}
\begin{eqnarray}
&& |\langle A_+ B_+^-|\hat{P}_+ | e_{B,+} \rangle|^2 =\left( 1-
p_{B\rightarrow A}^{-} \right) d_A^2, \label{eq:eexc-matr-a} \\
&& |\langle T_0^-|\hat{P}_- | e_{B,+} \rangle|^2 = \frac{1-
p_{B\rightarrow A}^{-}}{2}  d_A^2, \label{eq:eexc-matr-b} \\
&& |\langle S^-|\hat{P}_- | e_{B,+} \rangle|^2 = \frac{1+
p_{B\rightarrow A}^{-} - 2 p_{A\rightarrow B}^{-}}{2} d_A^2.
\label{eq:eexc-matr-c}
\end{eqnarray}
\end{subequations}
Inserting these matrix elements into the spectral representation
of $\theta_F(E)$, Eq.~(\ref{eq:frot-g}), we find for the FR angle
\begin{eqnarray}
&& \theta_F (E) = \frac{C E d_A^2}{2} \left[
\left(1- p_{B \rightarrow A}^{-} \right)
 \frac{E-E^-_{T_0 B}}{(E-E^-_{T_0 B})^2 + \Gamma^2}
\right. \hspace*{0.5cm} \nonumber
\\ && \hspace*{0.5cm} \left. -   \left(
1+ p_{B \rightarrow A}^{-} - 2 p_{A \rightarrow B}^{-}  \right)
 \frac{E-E^-_{SB}}{(E-E^-_{SB})^2 + \Gamma^2}  \right]
\label{eq:frot-5}
\end{eqnarray}
for probe energies $E \simeq E_X^{A(0)}$, in close analogy to
Eq.~(\ref{eq:frot-2}) for optical spin injection. The energy
differences are defined by $E^-_{T_0 B}=E_{T_0^{-}} - E^B$ and
$E^-_{SB}=E_{S^{-}} - E^B$. Because of the exchange splitting
between $|T_0^{-}\rangle$ and $|S^{-}\rangle$, $\theta_F(E)$ will
in general exhibit several peaks and lack point inversion
symmetry. The functional dependence on probe energy is determined
by the transfer probabilities and the energy differences $E^-_{T_0
B}$ and $E^-_{SB}$. For a more detailed analysis which takes into
account all finite transition matrix elements up to ${\cal
O}(t_c^2)$, see Appendix~\ref{ap:eexc}.

Experiments on doped QD's could provide valuable information
supplementing the experimental data obtained for optical pumping.
The main advantage over optical spin injection is that spin
decoherence times are expected to be substantially longer because
they are not limited by electron-hole recombination. Even more
importantly, FR measurements on doped coupled QD's can clarify
whether spin transfer occurs predominantly between the conduction
or valence band levels because, for $t_c = 0$ and $t_v \neq 0$,
$\theta_F (E) \simeq 0$ for probe energies $E \simeq E_X^{A(0)}$.

\section{Comparison with experiment}
\label{sec:exp}

In order to compare the results of Sec.~\ref{sec:exc-exc} with
experimental data from Ref.~\onlinecite{ouyang:03}, we first
provide numerical values for $\delta E_c$, $\delta E_v$, $U_A$,
and $U_B$. The energy level spectrum of CdSe QD's is well
established both experimentally and
theoretically.~\cite{ekimov:93,norris:94} The absorption energies
$E_X^{A(0)} = 2.41 \, {\rm eV}$ and $E_X^{B(0)} = 2.06 \, {\rm
eV}$ in Ref.~\onlinecite{ouyang:03} are consistent with $r_A
\simeq 2.0\, {\rm nm}$ and $r_B \simeq 3.5 \, {\rm nm}$, and we
will use these radii for the following calculations. From
Ref.~\onlinecite{ekimov:93}, $\delta E_c \simeq 0.30 \, {\rm eV}$
and $\delta E_v \simeq -0.10 \, {\rm eV}$.

From the bulk values for the static dielectric constant, $\epsilon
= 9.7$, and the band masses in the conduction and valence band,
$m_c/m_e = 0.12$ and $m_v/m_e = 0.45$, one obtains the exciton
radius $5.4 \, {\rm
nm}$.~\cite{reeber:76,wheeler:62,geick:66,fu:99} The exciton
radius is larger than $r_{A,B}$, and electrons and holes are
strongly confined in the QD's as assumed in Eq.~(\ref{eq:ham}).
The characteristic energy scale of the Coulomb interaction is
$U_\nu \simeq e^2/4 \pi \epsilon \epsilon_0 r_\nu$. For the given
values of $r_A$ and $r_B$,  $U_A = 0.07 \, {\rm eV}$ and $U_B =
0.04 \, {\rm eV}$.

The Hamiltonian Eq.~(\ref{eq:ham}) does not take into account
biexciton shifts, the exciton fine structure, and inter-dot
Coulomb interactions. For CdSe QD's with radii $1.5$--$4\,{\rm
nm}$, the biexciton shift is of order $0.01$--$0.02 \,{\rm eV}$
(Ref.~\onlinecite{efros:96}) and the characteristic energy
splitting between bright and dark excitons is smaller than $0.01
\, {\rm eV}$.~\cite{achermann:03} The characteristic energy scale
for inter-dot Coulomb interactions is $U_{AB} \simeq e^2/4 \pi
\epsilon_0 \epsilon (r_A + r_B) \leq 0.03 \, {\rm eV}$. However,
it is relevant only if neither of the two QD's is electrically
neutral. The most important effect of the inter-dot Coulomb
interaction is to lower the energy eigenvalues of
$|\tilde{T}_0\rangle$ and $|\tilde{S}\rangle$
[Eq.~(\ref{eq:2exc-4s})] by $U_{AB}$. All these energy scales are
small compared to the level broadening $\Gamma$ and can safely be
neglected.

In the following, we assume that only electrons in conduction band
levels are transferred between the QD's while valence band
electrons remain localized. As discussed in
Sec.~\ref{sec:doped-exc}, this assumption can be tested by
experiments on doped QD's. Mediated by electron transfer through
the molecular bridge, the lowest conduction band level in QD $B$
hybridizes with the lowest conduction band level in QD $A$.
Comparing the observed energy shift $E_X^{B}-E_X^{B(0)}=-0.02 \,
{\rm eV}$ with Eq.~(\ref{eq:1-exc-en}), we find
\begin{equation}
t_c =\sqrt{\left(E_X^{B(0)}-E_X^{B}\right)\left(\delta
E_c+U_B\right)} =0.082 \, {\rm eV}. \label{eq:transf-en}
\end{equation}

Our theory predicts that the exciton absorption peak for QD $A$ is
shifted to larger energies for the coupled QD's, in contrast to
the experimental result $E_X^{A}-E_X^{A(0)}<0$. The most likely
explanation for this is that the lowest conduction band level in
QD $A$ hybridizes also with higher excited levels in QD $B$ which
are nearly degenerate with $E_c^A$.~\cite{rem6} In order to
account for quantitative changes effected by this hybridization,
the energy $E_X^{A(0)}$ must be replaced by the true value of the
hybridized state in all expressions for the two-exciton
eigenenergies. This value can be obtained from $E_X^{A(0)} +
t_c^2/(\delta E_c+U_A)\simeq 2.36 \, {\rm eV}$, where the latter
is the experimental value for the exciton absorption edge of QD
$A$ in the coupled QD's. Hence, $E_{X}^{A(0)} \rightarrow 2.33 \,
{\rm eV}$.

From these parameters, we calculate for the transfer probabilities
between the lowest conduction band states $p_{A \rightarrow B} =
0.13$ and $p_{B \rightarrow A} = 0.06$. The energy differences
between the two-exciton states and the initial state are $E_{T_0
B} = 2.35 \, {\rm eV}$, $E_{SB}= 2.37 \, {\rm eV}$,
$E_{\tilde{T}_0 B} = 2.32 \, {\rm eV}$, $E_{\tilde{S}B}= 2.31 \,
{\rm eV}$. The oscillator strength for exciton creation,
proportional to $d_{A,B}^2$, is independent of the QD size in the
strong confinement regime and proportional to the QD volume for
weak confinement. Because both QD's are close to the strong
confinement limit, we assume a weak scaling $d_B^2/d_A^2 = 2$ for
the following Figures.

In Fig.~\ref{Fig5}(a), we show the FR angle calculated from
Eq.~(\ref{eq:frot-3}) as a function of probe energy for different
values of $\Gamma$, $\Gamma = 0.05 \, {\rm eV}$ (solid), $0.02 \,
{\rm eV}$ (dashed), and $0.08 \, {\rm eV}$ (dotted). We note that
even qualitative features depend strongly on the microscopic
parameters such as $\Gamma$. For small $\Gamma$, additional peaks
emerge because the contributions from the individual two-exciton
states can be resolved.

In spite of the dependence on microscopic parameters, some
pronounced features in $\theta_F (E)$ are generally present: (i)
$\theta_F (E)$ does not exhibit point-inversion symmetry, in stark
contrast to the FR angle expected from virtual transitions to a
single state. (ii) $\theta_F$ has in general more than two maxima
or minima. The positions and heights of the extrema are determined
by the interplay of the transfer probabilities $p_{A \rightarrow
B}$ and $p_{B \rightarrow A}$, and the energy splittings between
the different two-exciton states. Experiments have demonstrated
the strong dependence of the FR angle on the probe energy $E$,
including a fine structure of the resonance.~\cite{ouyang:03p}

In Fig.~\ref{Fig5}(b), we compare the calculated FR signal for
coupled QD's $A$ and $B$ with the corresponding result for
uncoupled QD's $A$ pumped at resonance. For a probe energy $E
\simeq 2.42 \, {\rm eV}$, the FR signal for coupled QD's $A$ and
$B$ is significantly smaller than the FR signal of the $AA$
system, consistent with experimental
observations.~\cite{ouyang:03}

Note that in Fig.~\ref{Fig5} we show the FR angle in arbitrary
units because its absolute numerical value depends sensitively on
unknown experimental parameters such as the packing density of
QD's and the number of QD's $A$ which are coupled to at least one
QD $B$. For a spin transfer probability $p_{B \rightarrow A} =
10$\%, assuming close packing of the QD's and that every QD $A$ is
coupled to one QD $B$, we estimate the maximum value of the
Faraday rotation angle to be $0.02$~rad for an $ABAABA$-structure
as investigated in Ref.~\onlinecite{ouyang:03}.

\begin{figure}
\centerline{\mbox{\includegraphics[width=7.3cm]{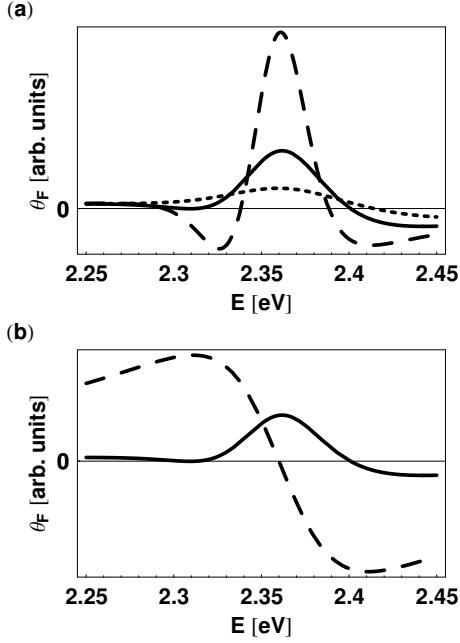}}}
\caption{(a) Plot of the FR angle as a function of probe pulse
frequency calculated from Eq.~(\ref{eq:frot-3}) for different
level broadenings $\Gamma = 0.05 \, {\rm eV}$ (solid), $0.02 \,
{\rm eV}$ (dashed), and $0.08 \, {\rm eV}$ (dotted). All other
parameters are as described in the text.  For small $\Gamma$,
$\theta_F(E)$ clearly shows the individual contributions from the
various two-exciton states. (b) Comparison of the FR angle for
coupled QD's for $\Gamma = 0.05 \, {\rm eV}$ (solid) with the
calculated signal for a $AA$ structure (dashed).} \label{Fig5}
\end{figure}

\begin{figure}
\centerline{\mbox{\includegraphics[width=8.3cm]{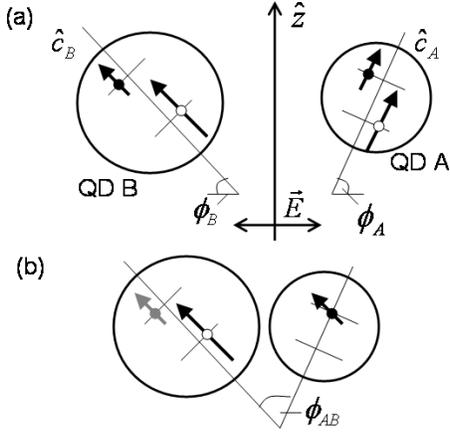}}}
\caption{(a) The hexagonal symmetry axes of QD's $A$ and $B$ are
in general oriented randomly relative to the direction of the
laser pump and probe pulse. Because of the interband selection
rules, a $\sigma^-$ circularly polarized laser pulse generates a
spin polarization along the symmetry axis of the respective QD.
(b) A conduction band electron created in QD $B$ retains its spin
direction on transfer to QD $A$. FR in QD $A$ probes the
projection of this spin onto the symmetry axis $\hat{\bf c}_A$,
which gives rise to a factor $\cos \phi_{AB}$ for the first and
second term in Eq.~(\ref{eq:frot-6}).} \label{Fig6}
\end{figure}

So far, we have assumed that the symmetry axis of the CdSe QD's
with hexagonal crystal structure is parallel to the propagation
direction of pump and probe laser pulses. However, in experiment
the QD's are randomly oriented. We discuss next how the random
orientation changes our results. The propagation direction of pump
and probe laser pulse is $\hat{\bf z}$, the polarization vector of
the probe pulse $\hat{\bf x}$, and the symmetry axes of QD's $A$
and $B$ are denoted by $\hat{\bf c}_A$ and $\hat{\bf c}_B$,
respectively. We define the azimuthal angles $\phi_{A} = \angle
(\hat{\bf x},\hat{\bf c}_{A})$ and $\phi_{B} = \angle (\hat{\bf
x},\hat{\bf c}_{B})$, and the angle enclosed by the two symmetry
axes $\phi_{AB} = \angle (\hat{\bf c}_{A},\hat{\bf c}_{B})$ [see
Fig.~\ref{Fig6}(a)]. The conduction band spin eigenstates with
quantization axis $\hat{\bf c}_{A,B}$ are denoted by $|\!
\uparrow_{A,B}\rangle$ and $|\! \downarrow_{A,B}\rangle$.

For arbitrary angle $\angle (\hat{\bf z},\hat{\bf c}_{B})$, the
probability for the circularly polarized pump pulse to create a
net spin polarization in the conduction band level decreases from
its maximum value at $\angle (\hat{\bf z},\hat{\bf c}_{B})=0$ to
zero at $\angle (\hat{\bf z},\hat{\bf c}_{B})=\pi/2$. For $\angle
(\hat{\bf z},\hat{\bf c}_{B}) < \pi/2$, the majority of conduction
band electrons is in spin state $|\! \uparrow_B\rangle$, with the
quantization axis defined by $\hat{\bf c}_{B}$. On transfer to QD
$A$, the conduction band electron retains its spin state because
states with $s_z = \pm 1/2$ are degenerate in both QD's and $t_c$
is spin-independent. The characteristic level spacing of valence
band states is large compared to the crystal field splitting in
bulk CdSe, which allows us to treat the latter as a small
perturbation, following Ref.~\onlinecite{efros:96}.

In the following, we calculate the FR angle for a random
orientation of QD's assuming that the pump pulse has created a
conduction band electron with spin $|\! \uparrow_B\rangle$. The
random orientation of QD's affects the FR of the probe pulse in
two ways. Firstly, the matrix elements for transitions from the
$j_z = \pm 3/2$ valence band levels to the $s_z = \pm 1/2$
conduction band levels in QD $A$ ($B$) decrease by $\sin \phi_A$
($\sin \phi_B$) compared to the oriented sample.~\cite{efros:96}
More importantly, also the relative orientation of $\hat{\bf
c}_{A}$ and $\hat{\bf c}_{B}$ modifies the FR angle. For
illustration, consider two QD's with $t_c=0$, and a conduction
band electron in spin state $|\! \uparrow_B\rangle$ in $B$. The
$\sigma^-$ circularly polarized component of the probe pulse with
$E \simeq E_X^A$ excites a virtual exciton in $A$, with a
conduction band electron in spin state $|\! \uparrow_A\rangle$.
Note that the spin direction is defined by $\hat{\bf c}_{A}$, the
symmetry axis of $A$. Expanding $|\! \uparrow_A\rangle =
\cos(\phi_{AB}/2)|\! \uparrow_B\rangle + i \sin(\phi_{AB}/2)|\!
\downarrow_B\rangle$ in terms of the eigenstates along
quantization axis $\hat{\bf c}_B$, the product state of the two
excitons contains terms in which the two conduction band spins are
antiparallel and have a finite overlap with the spin singlet
state. This is in stark contrast to the oriented sample, where the
two conduction band electrons would always form a triplet.

The analogous analysis for coupled QD's must take into account
both the reduced transition matrix elements for the probe pulse
and the relative orientation of QD's $A$ and $B$.  Because virtual
transitions to $|\tilde{T}_0 \rangle$ and $|\tilde{S} \rangle$
involve excitation of QD $B$ which was populated by the pump
pulse, the matrix elements in Eq.~(\ref{eq:2exc-matrb}) are
reduced by a factor $|\sin \phi_B|$ which is independent of the
relative orientation of $\hat{\bf c}_A$ and $\hat{\bf c}_B$. In
contrast, virtual transitions in QD $A$ probe the spin
polarization relative to the quantization axis $\hat{\bf c}_A$
after an electron with spin pointing along $\hat{\bf c}_B$ has
been transferred, and the transition matrix elements depend also
on $\phi_{AB}$ [Fig.~\ref{Fig6}(b)]. For the FR angle, we find
\begin{widetext}
\begin{eqnarray}
&& \theta_F (E) = \frac{CE}{2} \left\{ d_A^2  \cos \phi_{AB} \,
\sin^2 \phi_A \left[ \left(1- p_{B \rightarrow A} \right)
 \frac{E-E_{T_0 B}}{(E-E_{T_0 B})^2 + \Gamma^2} -
  \left(
1+ p_{B \rightarrow A} - 2 p_{A \rightarrow B}  \right)
 \frac{E-E_{SB}}{(E-E_{SB})^2 + \Gamma^2}  \right] \right.  \nonumber \\
&&  \hspace*{5cm} \left. - d_B^2 \sin^2 \phi_B \, p_{B \rightarrow
A} \left[
 \frac{E-E_{\tilde{T}_0 B}}{(E-E_{\tilde{T}_0 B})^2 + \Gamma^2} +
\frac{E-E_{\tilde{S}B}}{(E-E_{\tilde{S}B})^2 + \Gamma^2} \right]
\right\}. \label{eq:frot-6}
\end{eqnarray}
\end{widetext}

The dependence on the relative orientation of the two QD's,
$\phi_{AB}$, is readily understood. For $\phi_{AB} = \pi/2$, the
first and second term in the expression for $\theta_F(E)$ vanish
because the conduction band spin created in QD $B$ is
perpendicular to the spin quantization axis in QD $A$. A laser
pulse probing QD $A$ does not show any FR because the net spin
along $\hat{\bf c}_A$ vanishes [Fig.~\ref{Fig6}(b)].

In experiment, $\hat{\bf c}_A$ and $\hat{\bf c}_B$ are randomly
distributed over the unit sphere. Performing this average in
Eq.~(\ref{eq:frot-6}), we find for the FR angle
\begin{eqnarray}
 \overline{\theta_F} (E) &=& \frac{CE}{2} \left\{ \frac{3}{16}
d_A^2 \left[ \left(1- p_{B \rightarrow A} \right)
 \frac{E-E_{T_0 B}}{(E-E_{T_0 B})^2 + \Gamma^2}
\right. \right. \nonumber
\\ && \left.  -   \left(
1+ p_{B \rightarrow A} - 2 p_{A \rightarrow B}  \right)
 \frac{E-E_{SB}}{(E-E_{SB})^2 + \Gamma^2}  \right]  \nonumber \\
&&  - \frac{2}{3}d_B^2 p_{B \rightarrow A} \left[
 \frac{E-E_{\tilde{T}_0 B}}{(E-E_{\tilde{T}_0 B})^2 + \Gamma^2} \right. \nonumber \\
 && \left. \left. + \frac{E-E_{\tilde{S}B}}{(E-E_{\tilde{S}B})^2 + \Gamma^2} \right]
\right\}. \label{eq:frot-7}
\end{eqnarray}
Note that the spectral weight of the last term increases compared
to the oriented sample.

\section{Conclusion}
\label{sec:concl}

We have calculated the Faraday rotation angle for coupled QD's as
a function of the probe pulse frequency. We have considered an
initial spin polarization in neutral QD's (created by optical
pumping) and of one excess electron in the two coupled QD's. Our
results lead us to the following conclusions.

(i) The Faraday rotation angle shows a nontrivial functional
dependence on the probe energy, the details of which depend on the
spin exchange energy and spin transfer probabilities [see
Eq.~(\ref{eq:frot-3}) and Fig.~\ref{Fig5}(a)]. Most notably,
because several two-exciton states are separated in energy by a
small spin exchange coupling, $\theta_F(E)$ is not invariant under
point inversion symmetry. Measurement of $\theta_F(E)$ as a
function of probe energy would allow one to identify the
contributions of the various two-exciton states that are virtually
excited by the probe pulse.

(ii) Experiments on doped QD's would allow one to determine
whether spin transfer is mediated by transfer in the conduction or
valence band states. In particular, from a vanishing Faraday
rotation angle for probe pulse energies close to the resonance of
QD $A$ one could exclude that an excess electron injected into QD
$B$ has been transferred to $A$. In contrast, for optical spin
injection, spin could be transferred both between conduction and
valence band states.

(iii) In general, measurement of the Faraday rotation signal at a
given probe frequency does not provide enough information to
determine spin transfer probabilities between the QD's. However,
from the experimentally observed energy shifts, we calculate a
characteristic energy scale $t_c = 0.08 \, {\rm eV}$ for spin
transfer in the conduction band. Based on the transfer Hamiltonian
ansatz, this implies a probability of $6$\% for electron spin to
be transferred from QD $B$ to QD $A$, and of $13$\% for the
opposite direction.

The purpose of this work was to establish the connection between
spin transfer and the Faraday rotation signal observed in
experiment. Our analysis was based on a transfer Hamiltonian
ansatz. Some of the most interesting results of
Ref.~\onlinecite{ouyang:03} remain to be explored theoretically.
Most notably, the transfer Hamiltonian ansatz is based on the
assumption that electrons are transferred between the QD's via the
bridging benzene molecule. Microscopic work will have to clarify
why conjugated molecules provide efficient transfer paths between
QD's.

The results obtained here can provide important guidance also for
the identification of microscopic transfer mechanisms. The
increase of the Faraday rotation signal at a fixed probe frequency
has been interpreted as increase of the spin transfer efficiency
for higher temperatures.~\cite{ouyang:03} According to our
results, an increase in the transfer matrix element $t_c$ also
leads to a shift of the exciton edge in absorption spectra toward
lower energies. If the exciton absorption edge does not change
with increasing temperature, the increased Faraday rotation signal
is more likely effected, e.g., by additional incoherent transfer
paths than by an increase of the transfer matrix element.

\begin{acknowledgements}
We gratefully acknowledge helpful discussion with W. Lau, A.
Holleitner and, in particular, M. Ouyang. This work was supported
by DARPA, ONR, the EU TMR network MOLNANOMAG no.
HPRN-CT-1999-00012, the Swiss NCCR Nanoscience, and the Swiss NSF.
\end{acknowledgements}

\appendix

\section{Two-exciton eigenstates of coupled quantum dots}
\label{ap:2exc}

In order to evaluate the FR angle $\theta_F (E)$ from
Eq.~(\ref{eq:frot-g}) for arbitrary probe energies $E$, all
two-exciton intermediate states $|\psi_i \rangle$ with finite
transition matrix elements $\langle \psi_i | \hat{P}_{\pm}  |
X_{B,+} \rangle$ must be calculated. States with energies $E_i
\simeq E+E_X^{B(0)}$ lead to the dominant contributions in the
expression for the FR angle, Eq.~(\ref{eq:frot-g}). The states
$|A_+B_+\rangle$,  $|S\rangle$, and $|T_0\rangle$ defined in
Eqs.~(\ref{eq:2exc-1s}) and (\ref{eq:2exc-2s}) have energy
eigenvalues  $E_i$ with $|E_X^{A(0)}+E_X^{B(0)}-E_i |\leq {\cal
O}[t_c^2/(\delta E_c-U_A), t_c^2/(\delta E_c+U_A)]$, and are the
most important intermediate states for probe pulse energies $E
\simeq E_X^{A(0)}$. However, for the experimental values of
Ref.~\onlinecite{ouyang:03},  $ \delta E_v + U_A $ is small and
two additional two-exciton states must be taken into account.

The states
\begin{subequations}
\label{eq:2exc-4s}
\begin{eqnarray}
|\tilde{T}_0\rangle &=& \frac{1}{\sqrt{2}}
\left(\hat{c}_{c,-}^{A\dagger} \hat{c}_{c,+}^{B\dagger} +
\hat{c}_{c,+}^{A\dagger} \hat{c}_{c,-}^{B\dagger} \right)
\hat{c}_{v,+}^{B}  \hat{c}_{v,-}^{B}
|0 \rangle,\label{eq:2exc-4s-a} \\
|\tilde{S}\rangle &\propto & \frac{1}{\sqrt{2}}
\left(\hat{c}_{c,-}^{A\dagger} \hat{c}_{c,+}^{B\dagger} -
\hat{c}_{c,+}^{A\dagger} \hat{c}_{c,-}^{B\dagger} \right)
\hat{c}_{v,+}^{B}  \hat{c}_{v,-}^{B}
|0 \rangle  \nonumber\\
&& + \sqrt{2} \left(\frac{t_c}{\delta E_c + U_A + 2 U_B }
\hat{c}_{c,+}^{A\dagger}
\hat{c}_{c,-}^{A\dagger}  \label{eq:2exc-4s-b} \right. \\
&& \hspace*{1cm} \left. - \frac{t_c}{\delta E_c + U_B}
\hat{c}_{c,+}^{B\dagger} \hat{c}_{c,-}^{B\dagger} \right)
\hat{c}_{v,+}^{B}  \hat{c}_{v,-}^{B}  |0 \rangle \nonumber
\end{eqnarray}
\end{subequations}
differ from the corresponding states in Eq.~(\ref{eq:2exc-2s}) in
that both  holes are localized in QD $B$. The normalization
constant for $|\tilde{S}\rangle$ is fixed by $\langle \tilde{S}|
\tilde{S} \rangle = 1$. The eigenenergies
\begin{subequations}
\label{eq:2exc-4e}
\begin{eqnarray}
E_{\tilde{T}_0} &=&  E_X^{A(0)} +  E_X^{B(0)} + \delta E_v + U_A,
\label{eq:2exc-4e-a}  \\
E_{\tilde{S}} &=&  E_X^{A(0)} +  E_X^{B(0)} + \delta E_v + U_A
\label{eq:2exc-4e-b}\\ && \hspace*{1cm} + 2 t_c^2 \left(
\frac{1}{\delta E_c + U_B} - \frac{1}{\delta E_c + 2 U_B + U_A}
\right) \nonumber
\end{eqnarray}
\end{subequations}
are shifted relative to $E_{T_0}$ and $E_{S}$ by $\delta E_v+U_A$.

The state
\begin{eqnarray}
&& |\widetilde{B_+ B_-} \rangle \propto \left[
\hat{c}_{c,+}^{B\dagger} \hat{c}_{c,-}^{B\dagger} +
\frac{t_c}{\delta E_c + U_B} \left( \hat{c}_{c,-}^{A\dagger}
\hat{c}_{c,+}^{B\dagger} - \hat{c}_{c,+}^{A\dagger}
\hat{c}_{c,-}^{B\dagger}  \right) \right. \nonumber \\ && \left.
\hspace*{0.5cm} + \frac{2 t_c^2 \hat{c}_{c,+}^{A\dagger}
\hat{c}_{c,-}^{A\dagger}}{ (\delta E_c + U_B)(2 \delta E_c +U_A +
3 U_B)} \right] \hat{c}_{v,+}^{B}  \hat{c}_{v,-}^{B}  |0 \rangle
\label{eq:2exc-5s}
\end{eqnarray}
with
\begin{equation}
E_{ \widetilde{B_+ B_- }} = 2  E_X^{B(0)} - 2 \frac{t_c^2}{\delta
E_c + U_B} \label{eq:2exc-5e}
\end{equation}
is energetically separated from $E_{T_0}$ and $E_{S}$ by
$E_X^{A(0)}-E_X^{B(0)}$.

In Table~\ref{tab:2exc}, we summarize all two-exciton eigenstates
which contribute to the spectral representation of $\theta_F(E)$
up to order $t_c^2$. We also list the formal expressions for their
eigenenergies to leading order ${\cal O}(t_c^0)$ and give the
numerical values, taking into account terms up to ${\cal
O}(t_c^2)$ for the parameters discussed in Sec.~\ref{sec:exp}.

\begin{table}[t!]
\begin{tabular}{c|c|c}
$|\psi_i\rangle$ & $E_i$ & $E_i-E_X^{B}$ $[{\rm eV}]$ \\ \hline
$|A_+B_+ \rangle$ & $E_X^{A(0)}+E_X^{B(0)}$ & $2.35$ \\ \hline
$|T_0 \rangle$ & $E_X^{A(0)}+E_X^{B(0)}$ & $2.35$ \\ \hline $|S
\rangle$ & $E_X^{A(0)}+E_X^{B(0)}$ & $2.37$ \\ \hline $|B_+B_-
\rangle$ & $E_c^{B}-E_v^A+E_X^{B(0)}$ & $2.06$ \\ \hline
$|\tilde{T}_0 \rangle$ & $E_c^{A}-E_v^{B}+E_X^{B(0)}$ & $2.32$ \\
\hline
$|\tilde{S} \rangle$ & $E_c^{A}-E_v^{B}+E_X^{B(0)}$ & $2.33$ \\
\hline $|\widetilde{B_+B_-} \rangle$ & $2 E_X^{B(0)}$ & $2.04$ \\
\hline
\end{tabular}
\caption{Two-exciton eigenstates $|\psi_i\rangle$ which contribute
to the FR angle up to second order in $t_c$. We also list the
corresponding eigenenergies to ${\cal O}(t_c^0)$ and evaluate them
for the parameters discussed in Sec.~\ref{sec:exp}. As noted in
the main text, the degeneracy of $|\tilde{T}_0 \rangle$,
$|\tilde{S} \rangle$ with $|T_0 \rangle$, $|S\rangle$ is a
consequence of $\delta E_v + U_A \simeq 0$ for the QD's used in
experiment.} \label{tab:2exc}
\end{table}

\section{Eigenstates of doped coupled quantum dots}
\label{ap:eexc}

Here, we calculate eigenstates and energy eigenvalues for states
with two electrons and one hole in the coupled QD's. These are the
intermediate states $|\psi_i\rangle$ in Eq.~(\ref{eq:frot-g})
which have finite overlap matrix elements with
$\hat{P}_\pm|e_{B,+}\rangle$ and determine the FR angle for
coupled QD's doped with a single excess electron.

In addition to the states $|A_+B_+ ^-\rangle$, $|S^-\rangle$, and
$|T_0^-\rangle$ defined in Eqs.~(\ref{eq:eexc-1s}) and
(\ref{eq:eexc-2s}), five states have contributions of order
$t_c^2$ to the FR angle. These are
\begin{widetext}
\begin{subequations}
\label{eq:eexc-3s}
\begin{eqnarray}
|B_+B_- ^-\rangle &\propto&  \left[ \hat{c}_{c,+}^{B\dagger}
\hat{c}_{c,-}^{B\dagger}  + \frac{t_c}{\delta
E_c-U_A-U_B}\left(\hat{c}_{c,-}^{A\dagger}
\hat{c}_{c,+}^{B\dagger} - \hat{c}_{c,+}^{A\dagger}
\hat{c}_{c,-}^{B\dagger} \right) \right] \hat{c}_{v,-}^{A}
|0 \rangle, \label{eq:eexc-3s-a}   \\
|\widetilde{A_+B_+}^- \rangle & = & \hat{c}_{c,+}^{A\dagger}
\hat{c}_{c,+}^{B\dagger} \hat{c}_{v,+}^{B} |0 \rangle, \label{eq:eexc-3s-b}\\
|\tilde{T}_0^- \rangle &=& \frac{1}{\sqrt{2}}
\left(\hat{c}_{c,-}^{A\dagger} \hat{c}_{c,+}^{B\dagger} +
\hat{c}_{c,+}^{A\dagger} \hat{c}_{c,-}^{B\dagger} \right)
\hat{c}_{v,-}^{B}
|0 \rangle, \label{eq:eexc-3s-c} \\
|\tilde{S}^-\rangle &\propto  &  \left[ \frac{1}{\sqrt{2}}
\left(\hat{c}_{c,-}^{A\dagger} \hat{c}_{c,+}^{B\dagger} -
\hat{c}_{c,+}^{A\dagger} \hat{c}_{c,-}^{B\dagger} \right)
 + \sqrt{2} \left(\frac{t_c}{\delta
E_c + U_A +U_B} \hat{c}_{c,+}^{A\dagger} \hat{c}_{c,-}^{A\dagger}
- \frac{t_c}{\delta E_c } \hat{c}_{c,+}^{B\dagger}
\hat{c}_{c,-}^{B\dagger} \right) \right] \hat{c}_{v,-}^{B}  |0
\rangle,
\label{eq:eexc-3s-d}  \\
|\widetilde{B_+B_-}^- \rangle &\propto & \left[
\hat{c}_{c,+}^{B\dagger} \hat{c}_{c,-}^{B\dagger}
 + \frac{t_c}{\delta
E_c}\left(\hat{c}_{c,-}^{A\dagger} \hat{c}_{c,+}^{B\dagger} -
\hat{c}_{c,+}^{A\dagger} \hat{c}_{c,-}^{B\dagger} \right) \right]
\hat{c}_{v,-}^{B} |0 \rangle, \label{eq:eexc-3s-e}
\end{eqnarray}
\end{subequations}
\end{widetext}
with the proportionality constants chosen to ensure
normalization. The corresponding energy eigenvalues are
\begin{subequations}
\label{eq:eexc-3e}
\begin{eqnarray}
E_{B_+B_- ^- } &=& 2 E_c^B + U_B - E_v^A \label{eq:eexc-3e-a} \\
&& \hspace*{0.3cm} - 2
\frac{t_c^2}{\delta E_c - U_A - U_B},   \\
E_{\widetilde{A_+B_+}^-} &=& E_X^{B(0)}+E_c^A, \label{eq:eexc-3e-b}  \\
E_{\tilde{T}_0^-} &=&  E_X^{B(0)}+E_c^A, \label{eq:eexc-3e-c}  \\
E_{\tilde{S}^-} &=&  E_X^{B(0)}+E_c^A \label{eq:eexc-3e-d}
\\ && \hspace*{0.3cm} + 2 t_c^2 \left(\frac{1}{\delta
E_c}-\frac{1}{\delta E_c + U_A + U_B} \right),
\nonumber \\
E_{\widetilde{B_+B_-}^-} &=&  E_X^{B(0)}+E_c^B - 2
\frac{t_c^2}{\delta E_c}. \label{eq:eexc-3e-e}
\end{eqnarray}
\end{subequations}
From Eq.~(\ref{eq:eexc-3s}), we obtain the transition matrix
elements in terms of the transfer probabilities defined in
Eq.~(\ref{eq:tun-prob2}),
\begin{subequations}
\label{eq:eexc-matr2}
\begin{eqnarray}
|\langle B_+ B_- ^- |\hat{P}_- | e_{B,+} \rangle|^2 & = &
p_{A\rightarrow B}^{-} d_A^2, \label{eq:eexc-matr2-a} \\
|\langle \widetilde{A_+ B_+}^- |\hat{P}_+ | e_{B,+} \rangle|^2 & =
& p_{B\rightarrow A}^{-} d_B^2, \label{eq:eexc-matr2-b} \\
|\langle \tilde{T}_0^- |\hat{P}_- | e_{B,+} \rangle|^2 & = &
\frac{p_{B\rightarrow A}^{-}}{2} d_B^2, \label{eq:eexc-matr2-c} \\
|\langle \tilde{S}^- |\hat{P}_- | e_{B,+} \rangle|^2 & = & \frac{
p_{B\rightarrow A}^{-}}{2}  d_B^2, \label{eq:eexc-matr2-d} \\
|\langle \widetilde{B_+ B_-}^- |\hat{P}_- | e_{B,+} \rangle|^2 & =
& \left( 1 - p_{B\rightarrow A}^{-}\right) d_B^2.
\label{eq:eexc-matr2-e}
\end{eqnarray}
\end{subequations}
These transition matrix elements and the eigenenergies allow one
to calculate $\theta_F (E)$ for arbitrary energies. However, the
states in Eq.~(\ref{eq:eexc-3s}) are offset in energy from
$E_X^{A(0)} +E^B$. For probe energies $E \simeq E_X^{A(0)}$,
virtual transitions to the states $|A_+B_+ ^-\rangle$,
$|S^-\rangle$, and $|T_0^-\rangle$ are dominant, and $\theta_F
(E)$ simplifies to the approximate expression given in
Eq.~(\ref{eq:frot-5}).

\begin{table}[t!]
\vspace*{1cm}
\begin{tabular}{c|c|c}
$|\psi_i\rangle$ & $E_i$ & $E_i-E_c^B$ $[{\rm eV}]$\\ \hline
$|A_+B_+ ^- \rangle$ & $E_X^{A(0)}+E_c^B$ & $2.34$ \\ \hline
$|T_0^- \rangle$ & $E_X^{A(0)}+E_c^B$ & $2.34$ \\ \hline $|S^-
\rangle$ & $E_X^{A(0)}+E_c^B$ & $2.36$ \\ \hline $|B_+B_- ^-
\rangle$ & $2 E_c^{B}+U_B-E_v^A$ & $2.19$ \\ \hline
$|\widetilde{A_+B_-}^- \rangle$ & $E_X^{B(0)}+E_c^A$ & $2.27$ \\
\hline $|\tilde{T}_0^- \rangle$ & $E_X^{B(0)}+E_c^A$ & $2.27$ \\
\hline $|\tilde{S}^- \rangle$ & $E_X^{B(0)}+E_c^A$ & $2.28$ \\
\hline
$|\widetilde{B_+B_-}^- \rangle$ & $E_X^{B(0)}+E_c^B$ & $2.03$ \\
\hline
\end{tabular}
\caption{Eigenstates $|\psi_i\rangle$ with two electrons and one
hole which contribute to the FR angle up to second order in $t_c$.
We also list the corresponding eigenenergies to ${\cal O}(t_c^0)$
and evaluate them for the parameters discussed in
Sec.~\ref{sec:exp}. } \label{tab:eexc}
\end{table}

In Table~\ref{tab:eexc}, we list all states with two electrons and
one hole which contribute to $\theta_F(E)$ up to ${\cal
O}(t_c^2)$. We also provide the general expressions for the
eigenenergies to order ${\cal O}(t_c^0)$ and evaluate them
numerically for the parameters discussed in Sec.~\ref{sec:exp}.


\begin{thebibliography}{33}
\bibitem[{rem({\natexlab{a}})}]{rem1}
\bibinfo{note}{Note that a F{\"o}rster process, as discussed e.g. in A.~O.
  Govorov, Phys. Rev. B {\bf 68}, 075315 (2003), cannot explain the result of
  Ref.~\onlinecite{ouyang:03}, where spin is transferred from the larger to the
  smaller QD.}
\bibitem[{rem({\natexlab{b}})}]{rem2}
\bibinfo{note}{Two-exciton eigenstates which are energetically offset compared
  to $E+E_X^B$ will also contribute to the FR signal. However, their
  contribution varies slowly as a function of probe energy and leads at most to
  an offset in the results derived below. See also Appendix~\ref{ap:2exc}.}
\bibitem[{rem({\natexlab{c}})}]{rem3}
\bibinfo{note}{In order to calculate $\theta_F (E)$ for arbitrary energies $E$,
  also virtual transitions to $ |B_+ B_- \rangle$,$ |\widetilde{B_+ B_-}
  \rangle$, and the vacuum state $|0\rangle$ have to be taken into account. The
  corresponding expression is omitted here, but can be directly obtained from
  Eq.~(\ref{eq:frot-g}).}
\bibitem[{rem({\natexlab{e}})}]{rem6}
\bibinfo{note}{This hybridization with higher lying levels in QD $B$ does not
  invalidate our calculation of the FR angle in Sec.~\ref{sec:exc-exc} because
  the pump pulse leads to occupation of the lowest conduction band level in QD
  $B$ only.}
\end{thebibliography}
\end{document}